\documentclass[reprint,onecolumn,amsmath,amssymb,longbiobiography,preprint,superscriptaddress,floatfix]{revtex4-2}
\usepackage{CJK}
\usepackage{chemfig}
\usepackage{chemformula}
\usepackage{graphicx}
\usepackage{setspace}
\usepackage{subcaption}
\usepackage{grffile}
\usepackage{bm}
\usepackage{color}
\newcommand{\bw}{\begin{widetext}}
	\newcommand{\ew}{\end{widetext}}
\newcommand{\be}{\begin{equation}}
	\newcommand{\ee}{\end{equation}}
\newcommand{\baa}{\begin{eqnarray}}
	\newcommand{\ea}{\end{eqnarray}}

\newcommand{\1}{\mbox{\bf 1}}
\newcommand{\ud}{\mathrm{d}}

\newcommand{\vect}[1]{\bm{#1}}

\newcommand{\uns}[1]{\ensuremath{\unskip\,\mathrm{\scriptscriptstyle #1}}}
\newcommand{\un}[1]{\ensuremath{\unskip\,\mathrm{#1}}}

\begin{document}
	\begin{CJK*}{UTF8}{gbsn}
		\CJKfamily{gbsn}
		
\title{Water Absorption Dynamics in Medical Foam: Empirical Validation of the Lucas-Washburn Model}
\author{Weihua Mu(牟维华)}
\email{muwh@ucas.ac.cn}
\affiliation{Wenzhou Key Laboratory of Biomaterials and Engineering, Wenzhou Institute, University of Chinese Academy of Sciences, Wenzhou, Zhejiang 325000, China}
		
		\author{Lina Cao(曹丽娜)}
		\affiliation{Institute of Integrative Medicine, Department of Integrated Traditional Chinese and Western Medicine, Xiangya Hospital, Central South University, Changsha, 410008, P.R. China}
		\affiliation{Center for Interdisciplinary Research in Traditional Chinese Medicine, Xiangya Hospital, Central South University, Changsha, 410008, P.R. China}
		\affiliation{National Clinical Research Center for Geriatric Disorders, Xiangya Hospital, Central South University, Changsha, 410008, P.R. China}
		
\date{\today}

\begin{abstract}
This study extends the Lucas-Washburn theory through non-equilibrium thermodynamic analysis to examine fluid absorption in medical foams used for hemorrhage control. As a universal model for capillary flow in porous media, the theory demonstrated strong agreement with experimental results, confirming its semi-quantitative accuracy. Minor deviations, likely due to material heterogeneity, were observed and explained, enhancing the theory's applicability to real-world conditions. Our findings underscore the universality of the Lucas-Washburn framework and provide valuable insights for optimizing the design of medical foams, ultimately contributing to more effective bleeding control solutions in clinical applications.
\end{abstract}
\maketitle

\section{Introduction} 

Acute hemorrhage remains one of the leading causes of mortality in trauma care, particularly in cases involving non-compressible wounds where traditional hemostatic techniques, such as tourniquets, are ineffective. The development of advanced hemostatic materials that can rapidly absorb blood and promote clot formation is critical to improving outcomes in emergency and surgical settings. Hemostatic foams have emerged as a promising solution due to their ability to conform to irregular wound cavities and apply localized pressure to stop bleeding~\cite{blackbourne2011exsanguination,hickman2018biomaterials,feng2016chitosan,gao2020polymer,leonhardt2019absorbable,hong2019strongly,huang2021noncompressible,bao2022liquid,dong2022emerging,jiang2024superporous}.

In the use of medical foams, the kinetics of liquid absorption is crucial for their effectiveness, particularly in achieving rapid hemostasis. However, while significant research has been focused on material preparation and clinical evaluation, studies on the fundamental absorption kinetics are relatively scarce. This gap in the literature is critical, as the dynamics of fluid absorption, especially blood, are key to optimizing foam performance in emergency and surgical applications. Our work addresses this need by conducting a theoretical and experimental investigation of absorption kinetics using the well-established Lucas-Washburn rule within a non-equilibrium thermodynamic framework. By integrating material design with a detailed analysis of fluid absorption behavior, we aim to enhance the understanding of how medical foams function in real-world clinical scenarios, contributing to more effective and reliable hemostatic solutions.

The Lucas-Washburn equation, initially proposed over a century ago, has been a cornerstone in capillary flow studies. It has since been modified and extended to accommodate more complex systems, such as non-circular capillaries, heterogeneous geometries, and the influence of additional forces like gravity~\cite{cai2022capillary,lucas1918,washburn1921dynamics,blunt2013,higuera2014capillarity,dong2006imbibition,rye1996,cai2011,blunt1992simulation,Gorce2016}.
\begin{figure}[htbp]
\includegraphics[width=0.7\textwidth]{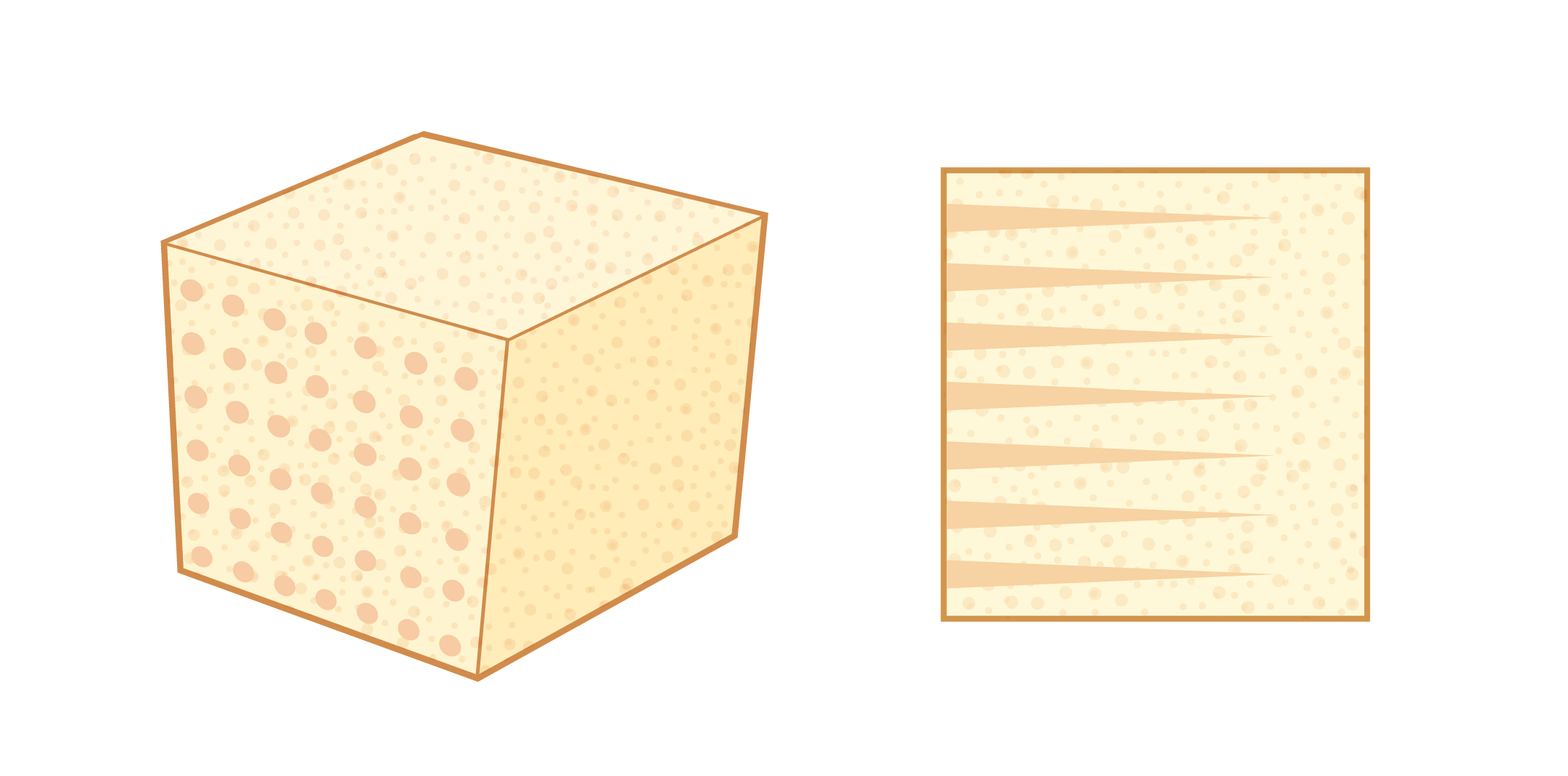}
\caption{The sketch illustrates the foam structure designed with a conical channel array. The foam itself is a porous material, with conical channels featuring a fixed length of $15,\un{mm}$ and varying radii of $6,,8,,10,\un{mm}$. These types of foams have also been investigated for their blood absorption kinetics~\cite{cao2023gelatin}}. 
\label{fig:sketch1}
\end{figure}

\section{Theoretical Modeling}

\begin{figure}[htbp]
	\includegraphics[width=0.7\textwidth]{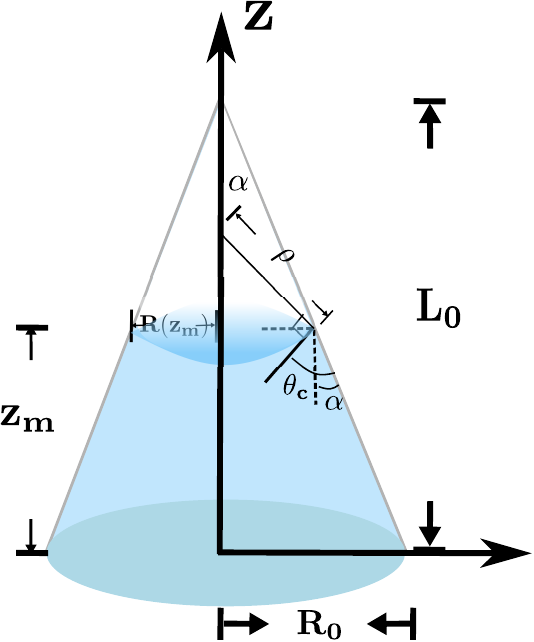}
	\caption{The sketch illustrates a conical channel, with the taper angle $\alpha$, the bottom radius $R_0$, the height of meniscus $z_m$, and the liquid-solid contact angle $\theta_c$ }. 
	\label{fig:channel}
\end{figure}

We investigate a water-absorbing sponge based on cellulose polymer materials, with artificially designed conical micro-channels embedded into the structure, providing an ideal example of capillary flow through a porous medium, as illustrated in Fig.~\ref{fig:sketch1}~\cite{cao2023gelatin}. In this study, for simplicity, we focus on homogeneous micro-channel structures, beginning with a representative channel shown in Fig.~\ref{fig:channel}, to analyze the liquid absorption behavior of the sponge. 
The volume of absorbed water is closely linked to the rising kinetics of the meniscus within the channel, which can be understood through the framework of non-equilibrium dissipation dynamics. Various approaches to modeling these kinetics have been developed since the work of Lucas (1918) and Washburn (1920), as discussed in Ref.~\cite{cai2022capillary} and the references therein.

We subsequently demonstrate how Onsager's principle~\cite{Doi2011}, can be applied to derive the time evolution of the meniscus position $z_{0}(t)$. This principle, fundamental in non-equilibrium thermodynamics, serves as a universal variational approach that allows for the derivation of kinetic equations for systems involving energy dissipation. By employing this minimization principle, one can efficiently obtain approximate solutions to these equations.

To Illustrate baisic physics by a simple example, we start with the nonequilibrium theory for the process towards equilibrium. For example, with the concerning of dissipation function $F_d=L_{ij}\dot{x}_i\dot{x}_j/2$, the Lagrangian equation for a system with kinetic energy $T(\dot{\vect{x}})$ and $V(\vect{x})$, $L=T-V$, are modified as~\cite{Landau1996}
\be\label{eom}
\dfrac{\ud}{\ud t}\dfrac{\partial{L(\dot{x}_i,\,x_i)}}{\partial \dot{x}_i}-\dfrac{\partial L}{\partial x_i} = -\dfrac{\partial F_d}{\partial \dot{x}_i},
\ee 
with $L_{ij}$ the kinetic coefficients for the nonequilibrium phenomena, 
and a general flux being $J_i\equiv L_{ij}\dot{x}_j$. Famous Onsager retroprcial relations shows $L_{ij}=L_{ji}$. For the case of zero-mass system, the first term of Eq. (\ref{eom}) can be neglected, leads to an overdampping approximation, 
\be\label{kinetic.eq}
\dfrac{\partial V(\vect{x})}{\partial x_i} + \dfrac{\partial F_d( \dot{\vect{x}})}{\partial \dot{x}_i}=0,
\ee  
For convenient, Rayleighian functional $R(\vect{x},\,\dot{\vect{x}})\equiv \partial_x V(\vect{x},\,\dot{\vect{x}})\dot{\vect{x}}+F_d$ is introduced, and Eq. (\ref{kinetic.eq}) has an equivalent compact form of $\partial R/\partial \dot{x}_i=0$, which is refered as Onsager's principle in literatures~\cite{Doi2011}.

For the system shown in Fig.~\ref{fig:sketch1}, to analyze the raising of the meniscus of the liquid in the converging conical tube, In the present case, consider the conical channel is a series of consequent straight tube pisewisely. The energy dissipation is therefore, 
\be\label{Fd}
F_d\left[\vect{v}(\vect{r})\right]=\dfrac{\eta}{2}\int\,d\vect{r}\,\left(\dfrac{\partial v_{\alpha}}{\partial x_{\beta}}+\dfrac{\partial v_{\beta}}{\partial x_{\alpha}}\right)^{2}.
\ee
Assuming a Poiseuille flow in a cylindrical tube, with $\vect{v}=v_{z}(r)\hat{e}_{z}$, and $v_{z}(r)=\left(R^{2}-r^{2}\right)\partial_z\, p/(4\eta)$, which is related to flux $Q$ by $\partial_{z}p=8\eta Q/\left(\rho\pi R^{4}(z)\right)$,
gets, 
\be\label{Fd1}
F_d\left[\vect{v}(\vect{r})\right]	\approx
\dfrac{4\eta Q^{2}}{\pi\alpha^{4}L_{0}^{3}\rho^{2}}\int_{0}^{\tilde{z}_{m}}\,\dfrac{d\tilde{z}}{\left(1-\tilde{z}\right)^{4}},
\ee
with dimensionless quantity $\tilde{z}\equiv z/L_0$ and $\tilde{z}_m\equiv z_{meni}/L_0$. The $L_0$ is the height of the conical tube, and $\alpha$ is the conical angle, thus the bottom radius of tube is $R_0=\alpha L_0$, and $R(z)=R_0-\alpha z=\alpha L_0\left(1-\tilde{z}\right)$. The flux $Q$ is related to the velocity of moving menisus through $Q=\pi R^2(z_{meni}) \dot{z}_{meni}$,
\be\label{Fd.zm}
F_d=4\eta L_{0}^{3}\left(1-\tilde{z}_{m}\right)^{4}\dot{\tilde{z}}_{m}^{2}\left(\tan^{-1}\tilde{z}_{m}+\dfrac{1}{2}\ln\dfrac{1+\tilde{z}_{m}}{1-\tilde{z}_{m}}\right).
\ee

Next, we formulate the energy change in the liquid-foam system, accounting for the work done by surface tension and gravitational potential energy, expressed as: \be\label{V}
V\approx2\pi\gamma\cos\theta_{c}\alpha L_{0}^{2}\int_{0}^{\tilde{z}_{m}}\,\left(1-\tilde{z}\right)d\tilde{z}-\pi\rho g\alpha^{2}L_{0}^{4}\int_{0}^{\tilde{z}_{m}}\tilde{z}d\tilde{z}\left(1-\tilde{z}\right)^{2},
\ee

The kinetic equation is obtained by Onsager principle, Eq.~(\ref{kinetic.eq}), gets, 
\bw
\be\label{long.eq.}
8\eta L_{0}^{3}\left(1-\tilde{z}_{m}\right)^{4}\dot{\tilde{z}}_{m}\left(\arctan\tilde{z}_{m}+\dfrac{1}{2}\ln\dfrac{1+\tilde{z}_{m}}{1-\tilde{z}_{m}}\right)+2\pi\gamma\cos\theta_{c}\alpha L_{0}^{2}\left(1-\tilde{z}_{m}\right)	\nonumber+\pi\rho g\alpha^{2}L_{0}^{4}\tilde{z}_{m}\left(1-\tilde{z}_{m}\right)^{2}=0.	
\ee
\ew
For the case of early stage of the kinectic process, $\tilde{z}_m\ll 1$, the kinetic equation is simplified to be 
\be\label{zdotz}
\tilde{z}_{m}\dot{\tilde{z}}_{m}\approx-\dfrac{\pi\alpha\gamma\cos\theta_{c}}{8\eta L_{0}}\left(1+\dfrac{\tilde{z}_m\,L_0^2}{2l_c^2\cos\theta_c}\right).
\ee 
Here, $l_c$ is the capillary length, and $l_c\sim 2.6\un{mm}$ for water. The length of conical channel is $L_0=15\,\un{mm}$, the contact angle is about $60^{\circ}$. for $\tilde{z}_m\ll 1$, approximately, 
\[
\tilde{z}_{m}\dot{\tilde{z}}_{m}\approx-\dfrac{\pi\alpha\gamma\cos\theta_{c}}{8\eta L_{0}},
\] 
which is a simple form of L-W formula.~\cite{cai2022capillary} 

The water is absorbed by the foam, filling both the voids in the porous material and the designed conical channels. The absorption capacity of the conical channels is significantly greater than that of the randomly distributed voids. Approximately, the volume of water absorbed by the foam can be described by:
\be\label{vol.ab}
V_{\uns{ab}}=\int_{0}^{z_{meni}}\,dz\,\pi\,R^{2}(z)=\alpha^{2}\pi L_{0}^{2}\left(\tilde{z}_{m}-\tilde{z}_{m}^{2}+\dfrac{\tilde{z}_{m}^{3}}{3}\right),
\ee
gets the time evolution for the weight of absorbed liquid, for the early stage, is roughly,
\[
M(t)\sim a\tilde{z}_{m}-b\tilde{z}_{m}^{2}\propto c_{1}t^{\nicefrac{1}{2}}-c_{2}t.
\]

\section{Experimental Verifcations and Discussions}

\begin{figure}[htbp]
	\includegraphics[width=0.8\textwidth]{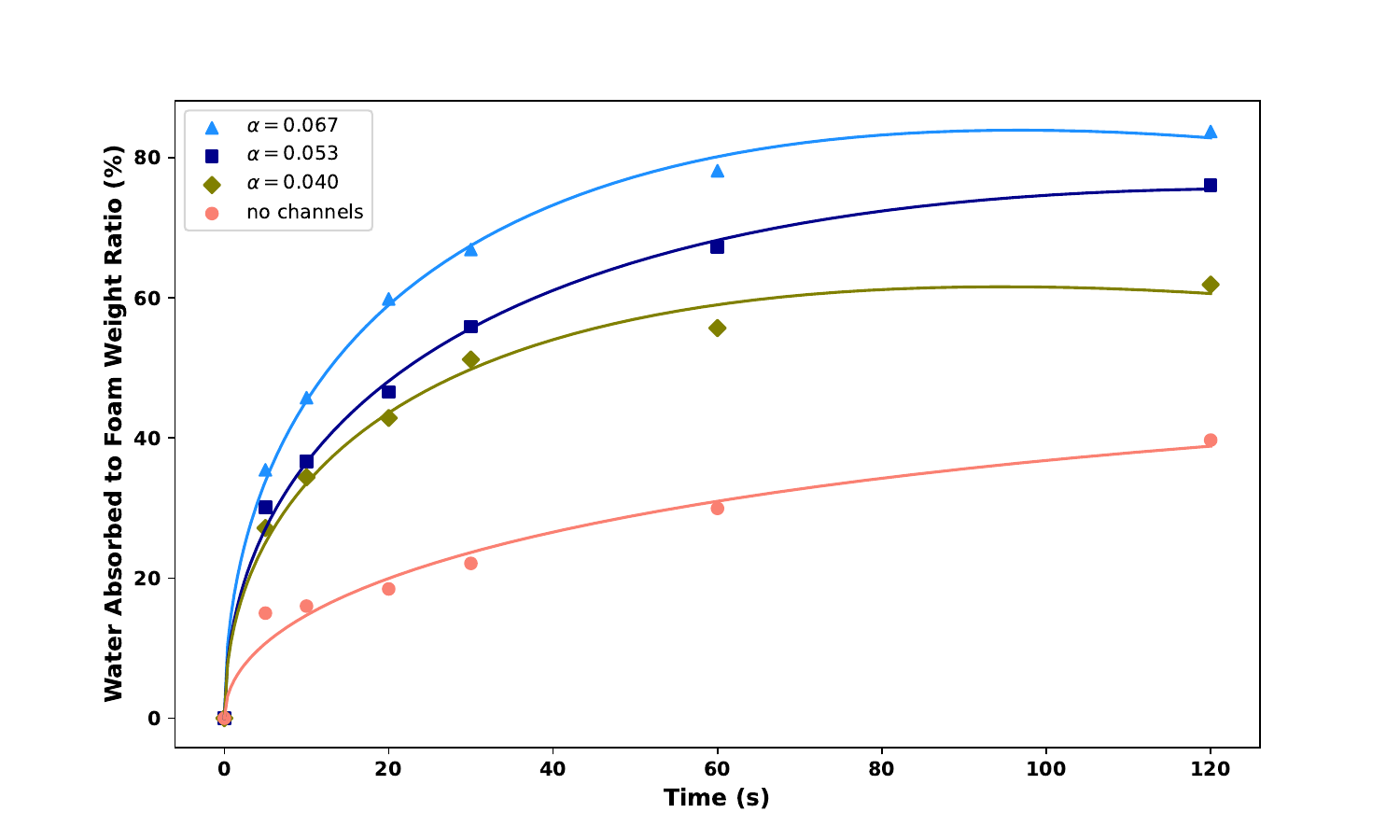}
	\caption{A comparative study of the theoretically predicted rule, $\Delta M/M_0 \sim a\,\sqrt{t} + b\,t$, and the experimental results is presented. The experiments involved a set of conical channels with conical angles $\alpha = 0.04,\, 0.053,\, 0.067$, alongside a channel-free sample used as a control for comparison. }
	\label{fig:Water_Absorption_Plot}
\end{figure}

Compared with the experimental results, we find that the semi-quantitative predictions are in good agreement with the observations, verifying the applicability of the Lucas-Washburn rule to the water absorption kinetics in this hemostatic foam system with artificially designed micro-channels, as shown in Fig.~\ref{fig:Water_Absorption_Plot}.

\begin{figure}[htbp]
\centering
\includegraphics[width=0.8\textwidth]{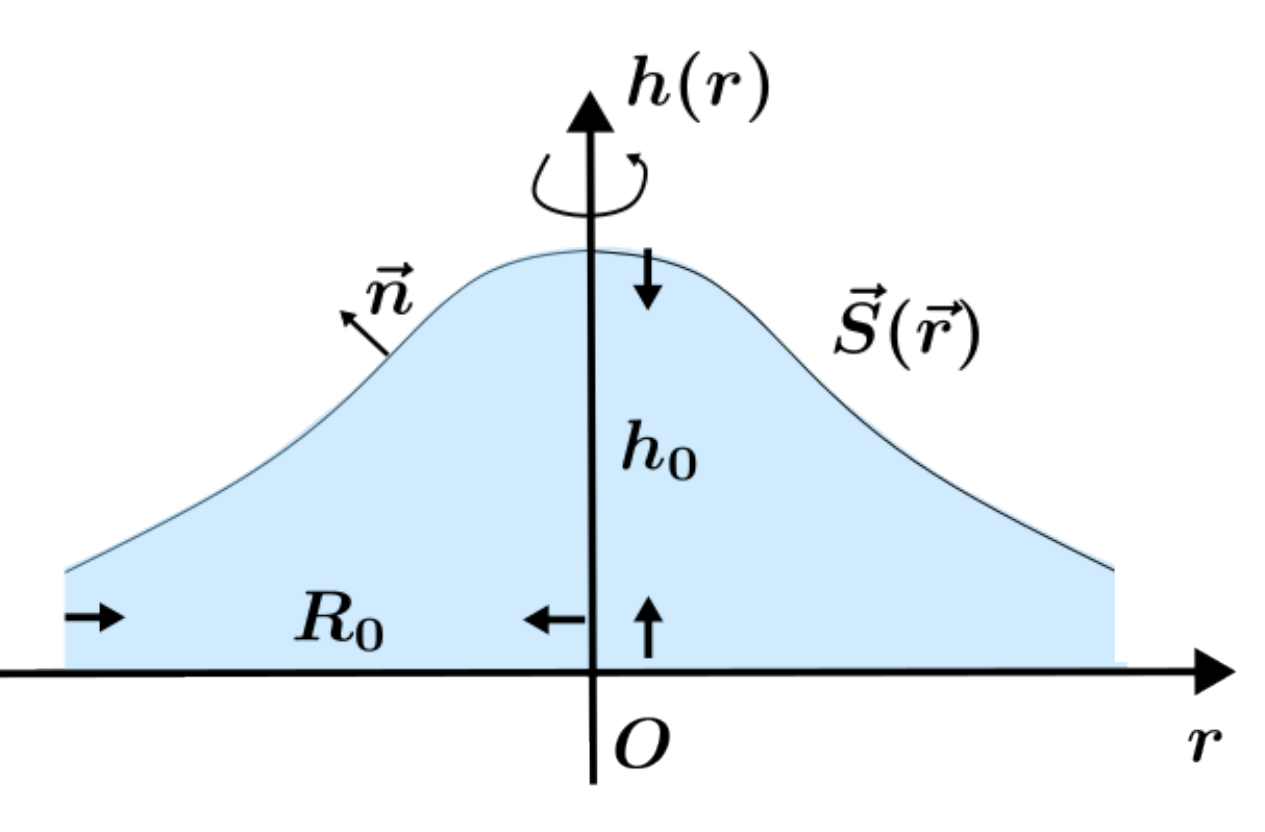}
\caption{A schematic representation of the channel-free foam material after water uptake (dyed blue) illustrates the formation of a waterfront. The waterfront forms an approximately convex surface of revolution, with a higher central region and a lower periphery. For a comparative reference, see the first row of Fig. 4A in Ref. ~\cite{cao2023gelatin}.}
\label{fig:waterfront.1023.2022.1}
\end{figure}

In our experiments, the water uptake behavior of the original channel-free foam material, which served as a control, exhibited a non-uniform waterfront profile. Specifically, this profile featured a higher central region and lower peripheries, as depicted in Fig.~\ref{fig:waterfront.1023.2022.1}.
This phenomenon can be attributed to the behavior of liquid-absorbing foam samples without channels, where the height of the liquid front at a given time, $t$, is proportional to $\sqrt{p^{\prime}_c}$. The capillary pressure, $p^{\prime}_c$, is inversely proportional to the characteristic void size, $d$, within the porous material, as described by the relationship $p^{\prime}_c \propto d^{-1}$. During the foam preparation process, the internal temperature is higher than the external environment, resulting in smaller voids in the core and larger ones near the surface. Assuming that the vertical accumulation of the absorbed liquid is faster than horizontal diffusion, the foam can be approximated as consisting of a series of independent, concentric cylindrical shells of water absorption (with radius $r$), each characterized by a uniform void size $d(r)$.

\begin{figure}[htbp]
	\centering
	\includegraphics[width=0.8\textwidth]{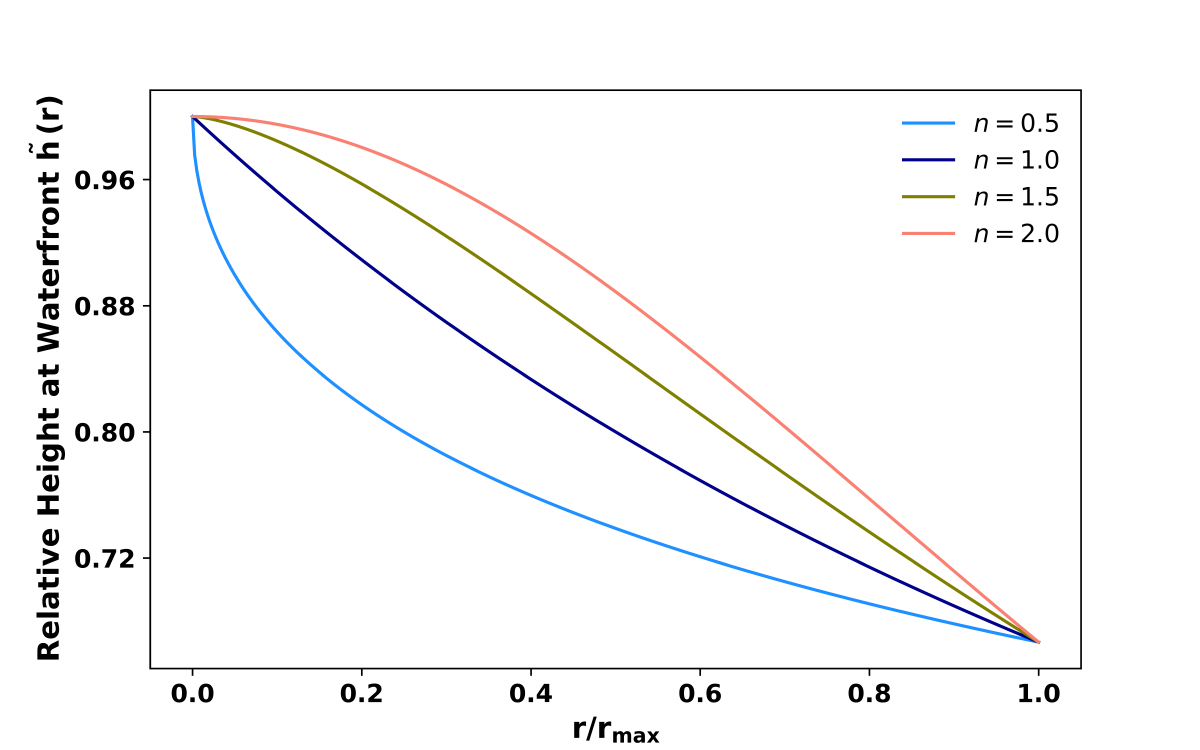}
	\caption{Assuming that the void size in the original channel-free foam material from the outer surface to the core follows a simple relationship given by $d(r)=d_0+\beta r^n$, the profile of the waterfront formed during water absorption in the foam sample can be accurately reproduced, as observed in the present experiment.}
	\label{fig:waterfront_profile}
\end{figure}

By proposing a distribution of void sizes following $d(r) = d_0 + \beta r^n$, the observed shape of the waterfront---higher in the center and lower at the edges, as seen in the channel-free control sample experiments---can be reproduced. For a rough estimation, assuming that the void size in the outermost layer is approximately $50\%$ smaller than in the core layer of the foam, suggests various waterfront profiles under different values of the parameter $\beta$, as shown in Fig.~\ref{fig:waterfront_profile}. This approach allows us to infer the spatial distribution of void sizes within the foam during its curing process by analyzing the shape of the foam's waterfront.

\section{Conclusion}

In this work, we employed non-equilibrium thermodynamic methods to derive the Lucas-Washburn theory, widely regarded as a fundamental model for fluid transport in channels and porous media. We focused on studying capillary flow based on the balance between viscous forces and surface tension, providing a universal framework applicable to a variety of systems.

Recent advancements have extended the Lucas-Washburn framework to more complex scenarios, such as non-Newtonian fluids, irregular porous geometries, and the influence of external fields like electric fields. However, challenges persist in accurately modeling these complex systems, where the assumptions underlying the traditional Lucas-Washburn equation may no longer hold (see Ref.~\onlinecite{cai2022capillary} and references therein). The universality of thermodynamic principles, such as Onsager's principle, suggests that modifications to the conservative force energy $V$ and dissipative energy $F_d$ could potentially extend the current theory to address these cases.

Our experimental investigation validated the theoretical predictions, showing that the Lucas-Washburn model accurately, and at least semi-quantitatively, describes the observed capillary flow behavior. We also identified deviations from the theoretical predictions, which we explained by considering factors such as the non-homogeneity of the medium---elements not accounted for in the original theory. Our refined analysis provided a comprehensive explanation for these discrepancies.

The Lucas-Washburn equation remains a cornerstone in capillary flow research, and future work will involve integrating advanced computational models and experimental techniques to overcome current limitations. There is growing interest in applying this theory to novel materials and emerging technologies. Continued refinement and adaptation of the Lucas-Washburn equation will enhance our understanding of capillary dynamics in increasingly complex systems.

\end{CJK*}
%



\end{document}